\documentclass[aps,prb,reprint,showpacs,superscriptaddress]{revtex4-1}

\usepackage{graphicx}
\usepackage{dcolumn}
\usepackage{bm}
\usepackage[dvipdfm]{hyperref}

\hypersetup{colorlinks=true,citecolor=blue,urlcolor=blue,linkcolor=blue}
\usepackage{subfigure}
\begin{document}



\title{Exploration of iron-chalcogenide superconductors}



\author{Chiheng Dong}
\affiliation {Department of Physics, Zhejiang University, Hangzhou 310027, China}

\author{Hangdong Wang}
\affiliation {Department of Physics, Zhejiang University, Hangzhou 310027, China}
\affiliation {Department of Physics, Hangzhou Normal University, Hangzhou 310016, China}

\author{Minghu Fang}
\email{mhfang@zju.edu.cn}
\affiliation {Department of Physics, Zhejiang University, Hangzhou 310027, China}


\date{\today}

\begin{abstract}
\noindent Iron-chalcogenide compounds with FeSe(Te,S) layers did not attract much attention until the discovery of high-$T_c$ superconductivity (SC) in the iron-pnictide compounds at the begining of 2008. Compared with FeAs-based superconductors, iron-chalcogenide superconductors have aroused enormous enthusiasm to study the relationship between SC and magnetisms with several distinct features, such as different antiferromagnetic ground states with relatively large moments in the parents, indicating possibly different superconducting mechanisms, the existence of the excess Fe atoms or Fe vacancies in the crystal lattice. Another reason is that the large single crystals are easily grown for the iron-chalcogenide compounds. This review will focus on our exploration for the iron-chalcogenide superconductors and discussion on several issues, including the crystal structure, magnetic properties, superconductivity, and phase separation. Some of them reach a consensus but some important questions still remain to be answered.


\end{abstract}


\maketitle
\section{Introduction}
The emergence of superconductivity (SC) with superconducting transition temperature (\textit{T}$_c$=26K) in the LaFeAs[O$_{1-x}$F$_x$] compounds \cite{LaOFFeAs}, following the same group's earlier discovery of superconductivity with \textit{T}$_c$ $\sim$ 5K in LaFePO$_{1-x}$F$_x$, is quite astonishing because the iron element is always considered to be ferromagnetic which is always detrimental to SC with spin singlet pairing. Soon after this discovery, Chinese scientists devoted great effort to pursuing new iron-based superconductors with new structures and higher \textit{T}$_c$. Significant progress was made in several months later. For example, various stackings of antifluorite Fe$_{2}$As$_{2}$ building blocks interleaved with alkali, alkaline earth, or rare earth oxides layers form varieties of Fe-As based superconductors, including '1111'  \cite{chenxh2008superconductivity,ChenGF2008Ce,CaoW56K}, '122'\cite{BaK,Co122,LJLiNi122}, '111'\cite{LiFeAs,TappLiFeAs} and '32522'\cite{WenHH32522}, '42622'\cite{Wen42622} system. The parent compound of FeAs-based superconductor undergoes a structural transition from tetragonal (T) to orthorhombic (O) lattice accompanied with an antiferromagnetic (AFM) transition occuring simultaneously or at a lower temperature ($T_N$) than the structural transition temperature ($T_S$) \cite{LaOFFeAsNeutron,CePhasediagram,Ba122Neutron}. The in-plane wave vector of AFM order in the FeAs-based parent compound is universally characterized by \textbf{Q}$_{AF}$=($\frac{1}{2}$, $\frac{1}{2}$)$_T$=(1, 0)$_O$ which is the same as the vector connecting the hole Fermi pocket at the zone center and the electron pocket at the corner of the Brillouin zone\cite{HongDBa122ARPES,Co122ARPES}. It was suggested that the FeAs-based material is an \textit{s}-wave superconductor with an unconventional spin fluctuation mechanism, that is, superconducting order parameter has opposite signs at the hole pocket and the electron pocket \cite{MazinPRL,KurokiTheorySwave,FaWTheory,BaKNeutron}.

The iron-chalcogenide compounds are much simpler in structure due to the neutrality of the FeSe(Te,S) layer than the Fe-As based compounds. The first discovery of SC with \textit{T}$_c$$\sim$8K in FeSe compound was reported by Hsu et al.\cite{maokun} on 15th July, 2008, and quickly followed by the reports of FeTe$_{1-x}$Se$_{x}$ ($T_c$$\sim$14K) by Fang et al.\cite{FangFeTeSe} on 30th July, 2008 and FeTe$_{1-x}$S$_x$ (\textit{T}$_c$$^{max}$$\sim$10K) \cite{fangFeTeS,FeTeS}. K$_{x}$Fe$_{2}$Se$_{2}$ with \textit{T}$_c$$\sim$~30K was firstly reported by Guo et al.\cite{xiaolong} on the 14th Dec., 2010 and considered to be of isostructure to the well-known ThCr$_2$Si$_2$ structure with I4/\textit{mmm} space group. On 23rd Dec., 2010, our group\cite{Fang122} independently reported on the bulk SC with \textit{T}$_c$$\sim$ 31K and a trace of SC at 40K in (Tl,K)Fe$_x$Se$_2$ system, and pointed out that the SC in this system is close to an AFM insulator, which is associated with the Fe-vacancy ordering in the Fe square lattice. Two kinds of Fe-vacancy orderings were suggested. Shortly after, SC with \textit{T}$_c$$\sim$30K in \textit{A}$_x$Fe$_{2-\delta}$Se$_2$ (\textit{A}=K\cite{GFChenFeSe122}, Rb\cite{XHRb122,WenHHFeSe122},Cs\cite{Cs}, and Tl/Rb\cite{TlRb}) compounds were reported. Electronic band structures and magnetic ground states of the \textit{A}$_x$Fe$_{2-\delta}$Se$_2$ compounds with Fe-vacancy orderings were systematically investigated\cite{CaoCTheory,TaoXiangTheory,YZhouTheory}, \textit{e.g.}, the first-principles calculations suggested that the ground state of (Tl,K)$_y$Fe$_{1.6}$Se$_2$ is of the checkerboard antiferromagnetically coupled blocks of the minimal Fe$_4$ square, which was confirmed by the neutron diffraction experiments later.

This review will focus on our exploration for the iron-chalcogenide superconductors and discussion on the several issues, including their crystal structures, magnetic properties, superconductivities, and phase separations. Some of them reach a consensus but some still remain to be answered.

\section{Fe(Te,Se,S) System}

\subsection{Discovery of the Superconductivity in Fe(Te,Se,S) system}

\begin{figure}
  \includegraphics[width=8.5cm]{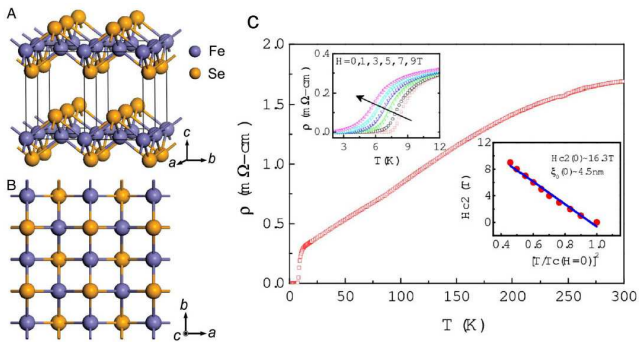}\\
  \caption{\label{Fig1}(color online) (a) Crystal structure of FeSe; (b) top view from the \textit{c} axis; (c) temperature dependence of electrical resistivity $\rho$ of FeSe$_{0.88}$. The left inset shows the $\rho$(\textit{T}) in the magnetic fields up to 9 T; the right inset displays the temperature dependence of upper critical field $H_{c2}$. From Ref. [21]}
\end{figure}

Iron selenium binary compounds have several phases with different crystal structures. SC occurs only in Fe$_{1+\delta}$Se with the lowest excess Fe\cite{FeSeSCvsExcess}, so-called the $\beta$ phase, which crystallizes into the anti-PbO tetragonal structure at ambient pressure (P4/\textit{nmm} ) \cite{maokun} and is considered to be the compound with the simplest structure in the Fe-based superconductors. The key ingredient of SC is a quasi-two-dimensional(2D) layer consisting of a square lattice of iron atoms with tetrahedrally coordinated bonds to the selenium anions which are staggered above and below the iron lattice [Fig.~\ref{Fig1}(a)].  These slabs, which are simply stacked and combined together with Van der Waals force, are believed to be responsible for the SC in this compound.

FeSe$_{0.88}$ was first found to be a superconductor with \textit{T}$_c$=8K by Hsu et al.\cite{maokun}, their results are summarized in Fig.~\ref{Fig1}(c). Then Medvedev et al.\cite{FeSepressure} found that the $T_c$ of Fe$_{1.01}$Se increases from 8.5 K to 36.7 K under an applied pressure of 8.9 Gpa. Although the structural transition from a tetragonal to hexagonal NiAs-type was observed above 7 Gpa, M\"{o}ssbauer measurements did not reveal any static magnetic order for the whole \textit{p}-\textit{T} phase diagram. However, short range spin fluctuations, which were strongly enhanced near \textit{T}$_c$, were observed by neutron magnetic resonance (NMR) measurements and thought to play an important role for the emergence of SC in this compound\cite{FeSeNMR}.

\begin{figure}
  \includegraphics[width=8.5cm]{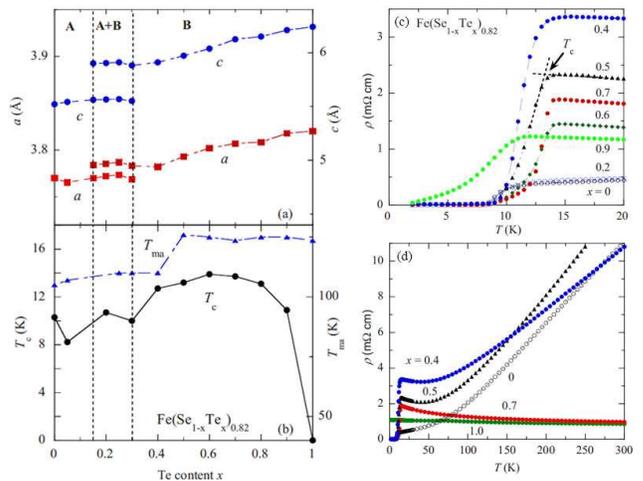}\\
  \caption{\label{Fig2}(color online)(a) Lattice parameters, (b) magnetic anomaly temperature \textit{T}$_{ma}$ and the onset superconducting transition temperature \textit{T}$_c$ each as a function of Te content \textit{x} in the Fe(Se$_{1-x}$Te$_x$)$_{0.82}$ series. (c) and (d) Temperature dependence of resistivity of Fe(Se$_{1-x}$Te$_x$)$_{0.82}$ with different values of \textit{x}. From Ref. [22]}
\end{figure}

\begin{figure}
  \includegraphics[width=8cm]{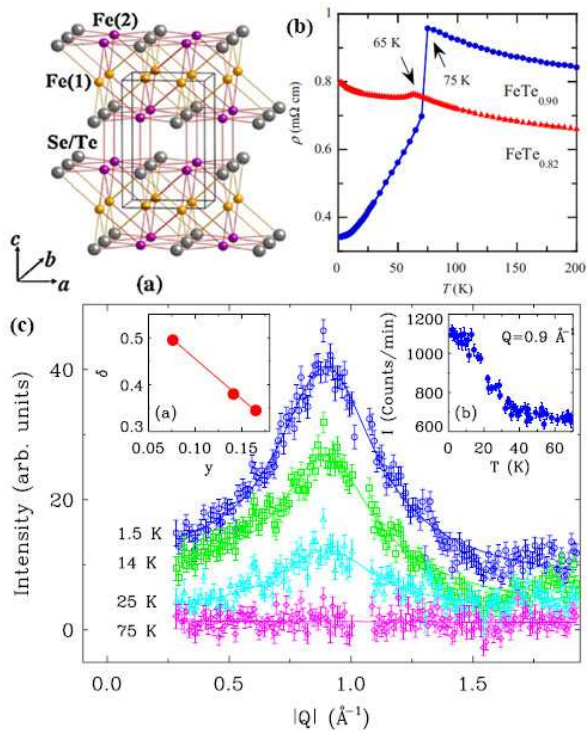}\\
  \caption{\label{Fig3}(color online)(a) Crystal structure of Fe$_{1+y}$Te. (b)Curves of resistivity as a function of temperature for FeTe$_{0.82}$ and FeTe$_{0.90}$. The arrows indicate the structural transition temperatures. (c) Short-range magnetic orders in the superconducting Fe$_{1.08}$Te$_{0.67}$Se$_{0.33}$ at different temperatures. The left inset shows the incommensurability $\delta$ as a function of \textit{y} for Fe$_{1+y}$Te. The right inset shows the intensity as a function of temperature.From Ref. [39]}

\end{figure}

Soon after the report of SC in FeSe by Hsu et al.\cite{maokun}, we first found\cite{FangFeTeSe} that the partial isovalent substitution of Te for Se in Fe(Te,Se) compounds results in the increase of the \textit{T}$_c$ to 14 K, and first reported that the end compound Fe$_{1+y}$Te is an AFM semiconductor or metal, which depends on the quantity of the excess Fe atoms in the lattice, instead of a superconductor, although the band calculation \cite{TheoryFeTe} indicated that it should be a superconductor with higher $T_C$ than that of Fe$_{1+y}$Se. Neutron diffraction experiments \cite{BaoweiFeTe} soon revealed that the AFM transition in Fe$_{1+y}$Te is also accompanied with a structural transition and that the AFM ground state at lower temperatures has two types, one is a commensurate AFM order in the sample with less excess Fe atoms, and the other is an incommensurate AFM order in the sample with more excess Fe atoms. It is important that the Fe$_{1+y}$Te be in a bicollinear AFM order, characterized by an in-plane propagation wave vector \textbf{Q}$_{AF}$=($\frac{1}{2}$, 0), which is different from the Fe-pnictide (1111 and 122 types) system. But for the superconducing Fe$_{1.08}$Te$_{0.67}$Se$_{0.33}$ sample, neutron scattering measurement revealed a prominent short-range quasi-elastic magnetic scattering at the incommensurate wave vector (0.438, 0). These short-range correlations are enhanced significantly as the temperature decreases below 40 K (Fig.~\ref{Fig3}(c)), thereby leading to an anomalous temperature dependence in the Hall coefficient. These results indicate strongly that the SC in Fe$_{1+y}$(Te,Se) compounds may be mediated by a magnetic fluctuation.

\subsection{Effect of excess Fe atoms on the structure and magnetism in Fe(Te,Se,S) system}

Although the Fe(Te,Se,S) system is considered to be the simplest structure in the iron-based superconductors, there are excess iron atoms partially occupying the interstitial sites between adjacent FeX (X = Te,Se,S) layers \cite{BaoweiFeTe} as denoted by Fe(2) in Fig.~\ref{Fig3}(a), which is similar to the location of Li in the Li$_{1-x}$FeAs superconductor that shares the same space group P4/\textit{nmm} with Fe$_{1+y}$Te \cite{TappLiFeAs}.

Zhang \textit{et al.}\cite{excessFetheory} studied the electronic and magnetic properties of the excess Fe atoms in Fe$_{1+x}$Te by means of density functional calculations. They found that the excess Fe atom has a monovalence in the Fe$_{1+x}$Te compound, \textit{i.e.} Fe$^{1+}$, and thus provides electron doping of approximately one \textit{e}/Fe. The excess Fe$^{1+}$ ion is suggested to be magnetic with 2.4 $\mu$$_B$ moment, bigger than that of the Fe$^{2+}$ (1.6-1.8$\mu$$_B$) cation at the Fe(1) position. The interaction between the local moment of the excess Fe$^{1+}$ ion and the moment of Fe$^{2+}$ on the layers is expected to persist even after the AFM order has been suppressed.

Generally, the excess Fe$^{1+}$ ions existing in both Fe$_{1+y}$Te and Fe$_{1+y}$Se, as well as Fe$_{1+y}$(Te,Se,S) lattices have an effect on their crystal structures and magnetic properties at low temperatures. McQueen \textit{et al.}\cite{FeSeStruvsExcessFe} found that Fe$_{1.01}$Se with less excess Fe atoms undergoes a structural transition at 90 K from a tetragonal structure to an orthorhombic structure, while there is no structural transition for Fe$_{1.03}$Se with more excess Fe atoms . The structure of Fe$_{1.01}$Se compound below 90 K can be identified to be orthorhombic (space group C\textit{mma}). The unit cell is enlarged into a $\sqrt{2}\times\sqrt{2}$ supercell similar to that of the iron-pnictide parent compound. There is no peak splitting or other significant change in the M\"{o}ssbauer spectrum corresponding to any magnetic transition between 50 and 100 K. However, the NMR measurement \cite{FeSeNMR} for the Fe$_{1.01}$Se compound demonstrated that the electronic properties of Fe$_{1.01}$Se are very similar to those of an optimally electron-doped FeAs-based superconductor and the AFM spin fluctuations are very strongly enhanced near \textit{T}$_{c}$.

As reported by Bao et al.\cite{BaoweiFeTe}, the parent compound Fe$_{1+y}$Te exhibits quite a different scenario. At the room temperature, Fe$_{1+y}$Te has a tetragonal structure (P4/\textit{nmm}), which is the same as that of Fe$_{1+y}$Se, regardless of the content of excess Fe atoms. But at lower temperatures, Fe$_{1+y}$Te compounds with different quantities of the excess Fe atoms undergo different structural transitions. For example, for Fe$_{1.141}$Te, a structural transition to an orthorhombic structure (P\textit{mmn}) occurs at \textit{T}$_s$$\approx$63 K, with the \textit{a} axis expanding and the \textit{b} axis contracting. While Fe$_{1.076}$Te with less excess atoms undergoes a first order transition to a monoclinic structure (space group P2$_{1}$/\textit{m}) at \textit{T}$_s$$\approx$75 K. For the low temperature phase, except for the difference between the \textit{a} and \textit{b} axis, the angle between \textit{a} and \textit{c} axes is less than 90$^{\circ}$. Unlike the structural transition in FeAs-based parent compounds, neither the orthorhombic nor monoclinic distortion in Fe$_{1+y}$Te compound doubles or rotates the the tetragonal cell.

\begin{figure}
  \includegraphics[width=8cm]{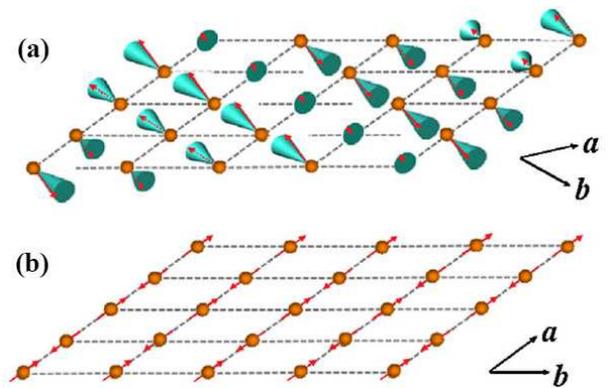}\\
  \caption{\label{Fig4}(color online) Magnetic structures of (a) FeTe; (b) SrFe$_2$As$_2$. From Ref. [39]}
\end{figure}

As discussed above, there is a discernable AFM transition coupled to the structural transition in Fe$_{1+y}$Te. The wave vector of AFM ordering observed in Fe$_{1+y}$Te is different from the ubiquitous in-plane wave vector of ($\frac{1}{2}$, $\frac{1}{2}$)$_{T}$ observed in the FeAs-based parent. Although the moments between the adjacent FeTe layers are antiferromagnetically aligned, the in-plane magnetic structure depends on the quantity of the excess Fe atoms. For example, Fe$_{1.068}$Te develops a bicollinear AFM order with an commensurate in-plane wave vector \textbf{Q}$_{AFM}$ = ($\frac{1}{2}$, 0) \cite{shiliangFeTeN,BaoweiFeTe}. The Fe ions in the plane have a total moment of 2.25(8) $\mu$$_B$ with a major component (2.0(7) $\mu$$_B$) along the \textit{b}-axis as shown in Fig.~\ref{Fig4}(a), which is rotated 45 $^{\circ}$ from \textit{a}-axis in the Fe-As materials. There are also projections of the moment along the \textit{a} and \textit{c} axes with -0.7(2) $\mu$$_B$ and 0.7(1) $\mu$$_B$ respectively. Li \textit{et al.}\cite{shiliangFeTeN} attribute the finite moments along the \textit{c}-axis to the finite moments of the excess Fe ion. As the content of the excess iron increases, the AFM order becomes incommensurate with an in-plane wave vector \textbf{Q}$_{AFM}$ = ($\pm\delta$, 0). The magnetic moments along the \textit{b}-axis are still ferromagnetically aligned. The row of the moments of Fe ions in the plane is modulated with the propagating vector 2$\pi\delta$/\textit{a}. Bao \textit{et al.}\cite{BaoweiFeTe} also found that incommensurability $\delta$ can be tuned by varying the excess Fe in the orthorhombic phase as shown in the left inset of Fig.~\ref{Fig3}(c). It reaches a commensurate value of 0.5 for the composition Fe$_{1.076}$Te.

For the superconducting FeSe$_{0.4}$Te$_{0.6}$ compound, neutron scattering measurements \cite{FeTeSeInelastic} revealed low energy spin fluctuations with a characteristic wave vector ($\frac{1}{2}$, $\frac{1}{2}$) which corresponds to Fermi surface nesting wave vector, but differs from the bi-collinear AFM order vector \textbf{Q}$_{AFM}$=($\frac{1}{2}$, 0) in the parent compound Fe$_{1+y}$Te. In addition, a pair of nearly compensated Fermi pockets \cite{FeTeARPES}, one hole pocket at the $\Gamma$ point of the Brillouin zone, the other electron pockets at the \textit{M} point were observed by the angle resolved photo emission spectroscopy (ARPES) experiments, which is similar to those expected from density functional theory (DFT) calculation\cite{TheoryFeSe}. However,no signature of Fermi surface nesting instability associated with Q$_{AFM}$=($\pi$/2, $\pi$/2) was observed, Fe$_{1+y}$(Te,Se) system may harbor an unusual mechanism for superconductivity.

\subsection{Effect of the excess Fe atoms on superconductivity in Fe(Te,Se,S) system}

\begin{figure}
  \includegraphics[width=8cm]{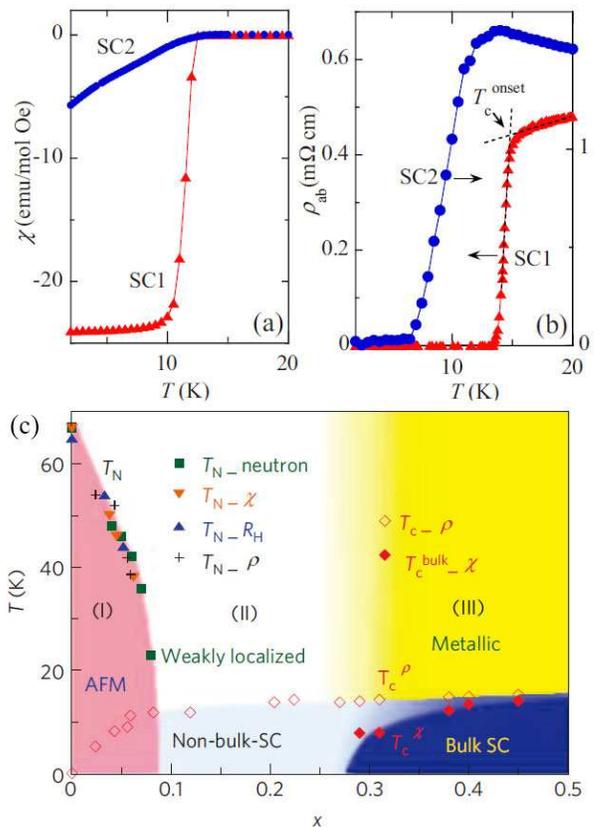}\\
  \caption{\label{Fig5}(color online) Temperature dependence of (a) susceptibility, (b) resistivity for Fe$_{1.03}$Te$_{0.63}$Se$_{0.37}$ (SC1) and Fe$_{1.11}$Te$_{0.64}$Se$_{0.36}$ (SC2). (c) Phase diagram of Fe$_{1.02}$Te$_{1-x}$Se$_{x}$. From Ref. [46,47]}
\end{figure}

\begin{figure}
  \includegraphics[width=8cm]{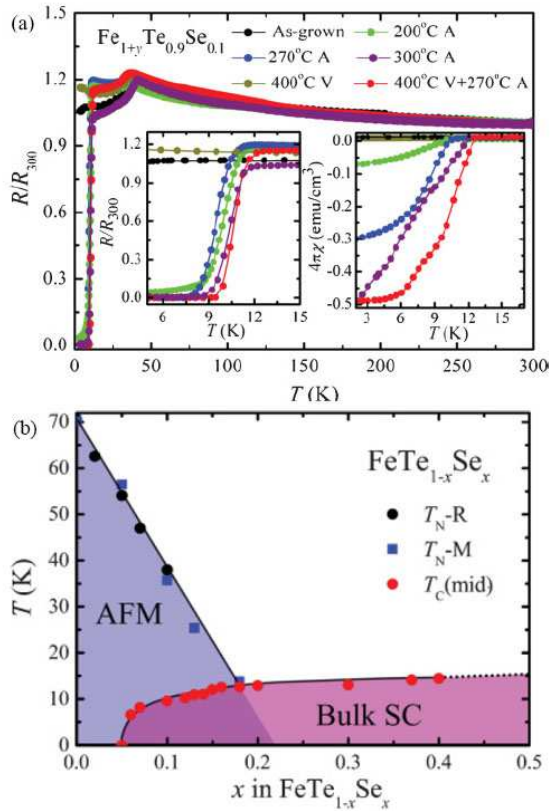}\\
  \caption{\label{Fig6}(color online) (a) Temperature dependences of the normalized in-plane resistivity for Fe$_{1+y}$Te$_{0.9}$Se$_{0.1}$ under different annealing conditions. Left inset shows the normalized \textit{R}(\textit{T}) near \textit{T}$_c$. Right inset shows the susceptibility as a function of temperature with H=5 Oe //\textit{ab}.(b) Phase diagram of Fe$_{1+y}$Te$_{1-x}$Se$_{x}$ with less excess irons.From Ref. [50]}
\end{figure}

The excess Fe atoms existing in Fe(Te,Se,S) lattice not only affect the crystal and magnetic structure, but also suppress their SC. Liu \textit{et al.} \cite{LTJExcess} studied the SCs in the optimally doped Fe$_{1+y}$Te$_{0.64}$Se$_{0.36}$ samples with different quantities of the excess Fe atoms. They found that Fe$_{1.11}$Te$_{0.64}$Se$_{0.36}$ (denoted as SC2) exhibits a lower superconducting volume fraction, a lower \textit{T}$_{c}$ and broader transition width as shown in Figs.~\ref{Fig5}(a) and (b), while Fe$_{1.03}$Te$_{0.64}$Se$_{0.36}$ (denoted as SC1) with less excess Fe atoms exhibits better SC. In addition, it was found that the resistivity in the normal state of SC1 shows a metallic behavior, whereas the resistivity of SC2 exhibits a logarithmic behavior, which is a characteristic of the weak localization of the charge carriers in two dimensions. They attributed these differences to the magnetic coupling between the excess Fe$^{1+}$ ions and the adjacent Fe$^{2+}$ on the square-planar sheets, which favors a short-range magnetic order. They further studied the phase diagram for the Fe$_{1.02}$Te$_{1-x}$Se$_{x}$ system, which can be divided into three composition regions with distinct physical properties (Fig.~\ref{Fig5}(c)) \cite{LTJFeTeSe}. The samples in Region 1 (0 $\leq$ \textit{x}$ \leq$0.09) exhibit a long-range AFM order with a wave vector ($\frac{1}{2}$, 0). The samples in Region 2 (0.09$\leq$\textit{x}$\leq$0.29) exhibit neither a long-range AFM order nor bulk superconductivity. Only the samples in Region 3 (\textit{x}$\geq$0.29) exhibit bulk superconductivity. They also suggested that the magnetic correlations near ($\frac{1}{2}$, 0) are antagonistic to SC and associated with a weak localization of charge carriers. Bulk superconductivity occurs only in the composition range where ($\frac{1}{2}$, 0) magnetic correlations are sufficiently suppressed and ($\frac{1}{2}$, $\frac{1}{2}$) spin fluctuation are associated with the nearly nesting Fermi surface dominates.

However, other group's conclusions for the phase diagram of Fe$_{1+y}$Te$_{1-x}$Se$_{x}$ system did not reach a consensus. For example, Katayama et al.\cite{SGFeTeSe} divided the phase diagram into three composition regions: the AFM phase for \textit{x}$\leq$0.1, SC emerges in \textit{x} $\geq$ 0.1, and the intermediate spin-glass region. Khasanov et al.\cite{uSRFeTeSe} suggested that in \textit{x}$\approx$0.25-0.45 region, SC coexists with an incommensurate AFM order, bulk SC did not appear until \textit{x}$\sim$0.5 . These discrepancies mainly concentrate on in what region bulk SC emerges and whether bulk SC with ($\frac{1}{2}$, $\frac{1}{2}$) spin resonance coexists with a long-range AFM order with the wave vector ($\frac{1}{2}$, 0). In fact, the excess Fe atoms existing unavoidably in the lattice suppress the SC and obscure the intrinsic properties. So an efficient way to remove these excess Fe is imperative.

Our group\cite{DongFeTeSe} found that the annealing in air can partially remove the excess Fe atoms in Fe$_{1+\delta}$Te$_{1-x}$Se$_{x}$ crystals, and can improve their SCs. As shown in Fig.~\ref{Fig6}(a), the as-grown Fe$_{1+y}$Te$_{0.9}$Se$_{0.1}$ crystal did not show any trace of SC above 2 K, but the crystal annealed in air for 2 h exhibits bulk SC. The AFM order transition at \textit{T}$_N$$\sim$38 K occurs also in the same crystal annealed in air, indicating the coexistence between bulk SC and AFM order in the crystals with a lower Se content. We also observed a similar enhancement of SC for the crystals with a higher Se content annealed in air .

In order to obtain the intrinsic phase diagram, we annealed all the Fe$_{1+y}$Te$_{1-x}$Se$_{x}$ (0.0$\leq\ x \leq$0.4) crystals in air at 270$^{\circ}$C for 2 h and obtained the phase diagram as shown in Fig.~\ref{Fig6}(b). We found that the AFM transition temperature \textit{T}$_{N}$ decreases with \textit{x}, and bulk SC emerges when \textit{x}$\geq$0.05. Obviously, there is a region where the bulk SC with ($\frac{1}{2}$, $\frac{1}{2}$) spin resonance coexists with the long-range AFM order with the wave vector \textbf{Q}$_{AFM}$=($\frac{1}{2}$, 0). Our finding of the similarity between the phase diagram of FeTe$_{1-x}$Se$_{x}$ system and those of (Ba,K)Fe$_2$As$_2$, BaFe$_{2-x}$Co$_x$As$_2$, as well as SmFeAsO$_{1-x}$F$_{x}$ system, indicates that the coexistence between the long-range AFM order and the bulk SC may be an intrinsic property of the Fe-based superconductors.

\section{(Tl,K,Rb)Fe$_x$Se$_2$ system}

\subsection{Discovery of superconductivity in (Tl,Rb,K)Fe$_x$Se$_{2}$ system}

\begin{figure}
  \includegraphics[width=8cm]{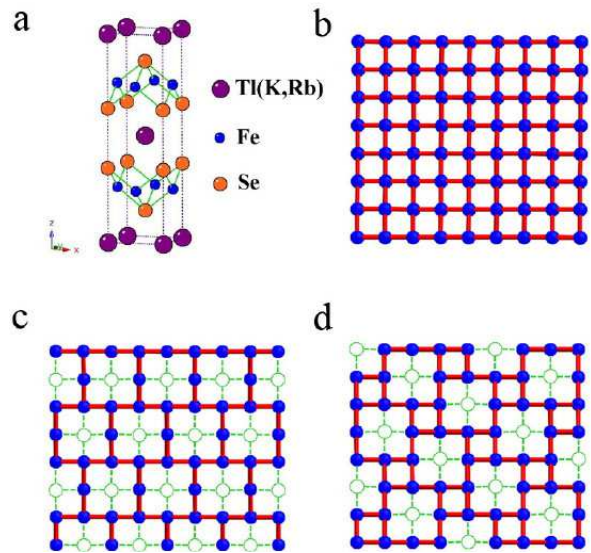}\\
  \caption{\label{Fig7}(color online) (a) Crystal structure of (Tl,K,Rb)Fe$_{x}$Se$_{2}$. (b) Fe$^{2+}$ square plane for stoichiometric (Tl,K,Rb)Fe$_{2}$Se$_{2}$, which share the common layer with all the other iron-based superconductors. Fe-vacancy ordering patterns for (c) (Tl,K,Rb)Fe$_{1.5}$Se$_{2}$ and (d)(Tl,K,Rb)Fe$_{1.6}$Se$_{2}$. From Ref. [26]}
\end{figure}

In order to develop an unified picture for high temperature superconductivity (HTSC) in both the Fe-based and cuprates compounds, a key strategy is to explore the possibility to tune the Fe-based compound into an insulator. Our first effort for this purpose is to study the physical properties of La$_2$O$_3$Fe$_2$(Se,S)$_2$ compounds \cite{LaOFeSe} and their doping compounds. We showed theoretically that they contain enhanced correlation effects through band narrowing compared with LaOFeAs, considering an Fe square lattice with an expanded unit cell, and provided experimental evidence that they are Mott AFM insulators with moderate charge gaps. But we found that it is difficult to let SC emerge in their doping compounds. Our second effort is to study the TlFe$_{x}$Se$_{2}$ compounds as early as 2009. As is well known, all the Fe-based superconductors share a common layered structure based on a square planar Fe$^{2+}$ layer as shown in Fig.~\ref{Fig7}(b). The ThCr$_{2}$Si$_{2}$-type crystal structure of the stoichiometric TlFe$_{2}$Se$_{2}$, which is shown in Fig.~\ref{Fig7}(a), is the same as that of BaFe$_{2}$As$_{2}$. However, Tl has a monovalence in the TlFe$_{2}$Se$_{2}$ compound and there are always vacancies on the Fe-square lattice. Therefore, the actual chemical formula should be TlFe$_{x}$Se$_{2}$ with 1.3$\leq$\textit{x}$<$2.0. There exists evidence that the Fe-ions in TlFe$_{x}$Se$_{2}$ with \textit{x}=1.5 exhibit long-range AFM order. According to the M\"{o}ssbauer and neutron diffraction studies, Haggstrom et al. \cite{TlFe2-xSe2Moss} and Sabrowsky et al.\cite{TlFe2-xSe2Neutron} argued that the AFM orders in TlFe$_{x}$Se$_{2}$ and TlFe$_{x}$S$_{2}$ are associated with the crystallization of the Fe-vacancies, as shown in Figs.~\ref{Fig7}(c) and 7(d) for \textit{x}=1.5 and 1.6, respectively. Soon after we reported that the suggestion about the Fe-vacancy super-lattice, the $\sqrt{5}\times\sqrt{5}$ super-lattice as shown in Fig.~\ref{Fig7}(d) was confirmed to be a common structure of \textit{A}$_{2}$Fe$_{4}$Se$_{5}$ (\textit{A}=K, Rb, Cs, Tl) by means of the neutron diffraction experments \cite{Baowei122,YeFNeutron122}. From this point of view, we expect the AFM order to be suppressed as Fe-content increases. The possible emergence of the SC in this new 122-type iron-chalcogenide motivated us to grow TlFe$_{x}$Se$_{2}$ single crystals with higher Fe-content.

\begin{figure}
  \includegraphics[width=8cm]{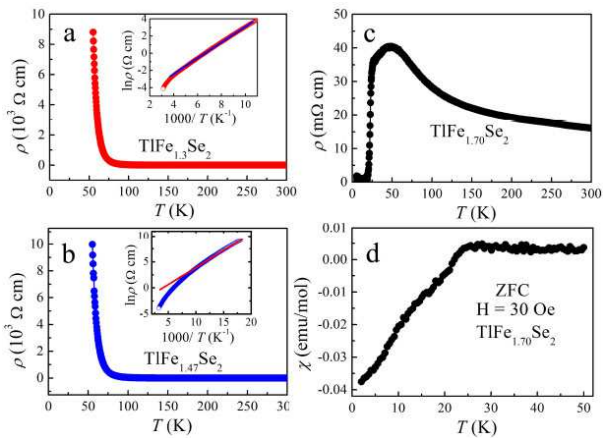}\\
  \caption{\label{Fig8}(color online) Temperature dependences of the in-plane resistivity $\rho$, and ln$\rho$ on 1/\textit{T} (in the inset) for both insulators. (a) TlFe$_{1.30}$Se$_{2}$ crystal. (b) TlFe$_{1.47}$Se$_{2}$ crystal. (c) Temperature dependence of the inplane resistivity, $\rho$(\textit{T}). (d) Magnetic susceptibility, $\chi$(\textit{T}), measured at 30 Oe in the zero-field-cooling (ZFC) process, for the TlFe$_{1.70}$Se$_{2}$ crystal. From Ref. [26]}
\end{figure}

At the beginning, in order to obtain the single crystals with higher Fe content, we tried to grow the TlFe$_{x}$Se$_{2}$ single crystals using various nominal compositions as the staring materials, \textit{i.e.}, Tl$_{z}$Fe$_{2}$Se$_{2}$ (0.4$\leq$\textit{z}$\leq$1.0), TlFe$_{z}$Se$_{2}$ (1.5$\leq$\textit{z}$\leq$3.0). We have obtained a series of TlFe$_{x}$Se$_{2}$ (1.3$\leq$\textit{x}$\leq$1.7) single crystals. The energy dispersive x-ray spectrometer (EDXS) indicates that the quantities of Tl and Se are respectively nearly 1.0 and 2.0, but the Fe content is always less than 2.0. The temperature dependences of resistivity for the crystals with less Fe content exhibit a thermally activated semiconducting behavior with activation energies of 80.2 meV and 57.7 meV for \textit{x}=1.30 and 1.47 respectively as shown by the fitting line in the inset of Fig.~\ref{Fig8}(a) and 8(b). SC with a transition temperature \textit{T}$_{c}$$^{mid}$=22.4 K and zero resistivity at \textit{T}$_{c}$$^{zero}$=20 K are indeed first observed in the TlFe$_{1.70}$Se$_{2}$ crystal, as showed in Fig.~\ref{Fig8}(c), while its superconducting volume fraction is very small ($<$1\%), see Fig.~\ref{Fig8}(d). It was first found that the evolution from an insulator to a superconductor in TlFe$_{x}$Se$_{2}$ system with the increase of Fe content. At the same time, we found that it is difficult to grow a TlFe$_{x}$Se$_{2}$ single crystal with \textit{x}$>$1.70.

\begin{figure}
  \includegraphics[width=8cm]{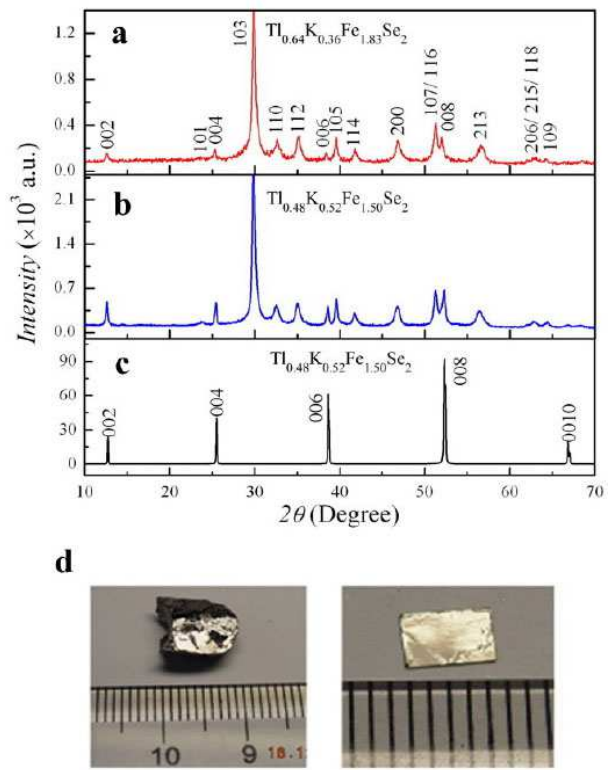}\\
  \caption{\label{Fig9}(color online) Powder XRD patterns for (a) Tl$_{0.64}$K$_{0.36}$Fe$_{1.83}$Se$_{2}$ and for (b) and (c) Tl$_{0.48}$K$_{0.52}$Fe$_{1.50}$Se$_{2}$. (d) Photos of Tl$_{0.48}$K$_{0.52}$Fe$_{1.50}$Se$_{2}$ single crystals. From Ref. [26]}
\end{figure}

According to the results mentioned above, it is necessary to further increase the Fe content to achieve bulk SC in this system. We tried to grow the Tl$_{1-y}$K$_{y}$Fe$_{x}$Se$_{2}$ single crystals by adding K in the starting materials with the nominal composition: Tl$_{0.5}$K$_{z}$Fe$_{2}$Se$_{2}$ (0.15$\leq$\textit{z}$\leq$0.45), Tl$_{0.4}$K$_{z}$Fe$_{2}$Se$_{2}$ (0.2$\leq$\textit{z}$\leq$0.5) and Tl$_{0.8-z}$K$_{z}$Fe$_{2}$Se$_{2}$ (0.1$\leq$\textit{z}$\leq$0.4). We finally obtained a series of Tl$_{1-z}$K$_{z}$Fe$_{x}$Se$_{2}$ (1.50$\leq$\textit{x}$\leq$1.88, 0.14$\leq$\textit{z}$\leq$0.57) single crystals as shown in Fig.~\ref{Fig9}(d). Figures~\ref{Fig8}(a)-(8)(c) show the single crystal and powder x-ray diffraction (XRD) patterns for Tl$_{0.64}$K$_{0.36}$Fe$_{1.83}$Se$_{2}$ and Tl$_{0.48}$K$_{0.52}$Fe$_{1.50}$Se$_{2}$. All the diffraction peaks can be well indexed with the tetragonal space group I4/\textit{mmm}. Lattice constants can be derived by fitting the powder XRD data: \textit{a}=3.88${\AA}$, \textit{c}=14.05${\AA}$ for Tl$_{0.64}$K$_{0.36}$Fe$_{1.83}$Se$_{2}$ and \textit{a}=3.90${\AA}$, \textit{c}=13.99${\AA}$ for Tl$_{0.48}$K$_{0.52}$Fe$_{1.50}$Se$_{2}$. Broader peaks in the Powder XRD patterns might indicate the existence of some disorder in the crystals, and it is difficult to observe the super-lattice diffraction peaks of the Fe vacancy order in our experiments. Only (00\textit{l}) peaks were observed in the single crystal XRD patterns, indicating that the crystallographic \textit{c} axis is perpendicular to the plane of the single crystal.

\begin{figure}
  \includegraphics[width=8.3cm]{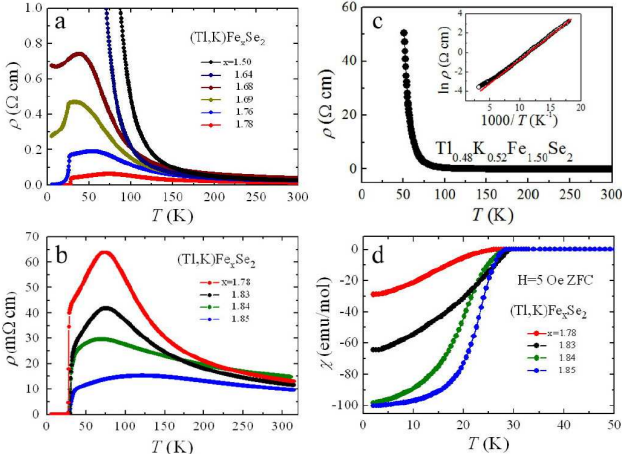}\\
  \caption{\label{Fig10}(color online) Temperature dependences of the resistivity and susceptibility for the (Tl,K)Fe$_{x}$Se$_{2}$ (1.5$\leq$\textit{z}$\leq$1.85) crystals. (a) In-plane resistivity, $\rho$(\textit{T}), for (Tl,K)Fe$_{x}$Se$_{2}$ (1.5$\leq$\textit{z}$\leq$1.78). (b) $\rho$(\textit{T}) for (Tl,K)Fe$_{x}$Se$_{2}$ (1.78$\leq$\textit{z}$\leq$1.85). (c) $\rho$(\textit{T}) for the Tl$_{0.48}$K$_{0.52}$Fe$_{1.50}$Se$_{2}$ crystal. Inset: ln$\rho$ as a function of 1/T. (d) $\chi$(\textit{T}) measured with 5 Oe, ZFC process. From Ref. [26]}
\end{figure}

Figure~\ref{Fig10} shows the curves of $\rho$(\textit{T}) and $\chi$(\textit{T}) for the (Tl,K)Fe$_{x}$Se$_{2}$ single crystals. The values of resistivity at the room temperature are 27.8 m$\Omega$ and 27.2 m$\Omega$ for \textit{x}=1.50 and 1.64 respectively. The resistivity increases rapidly as the temperature decreases (Fig.~\ref{Fig10}(c)), showing a thermally activated behavior: $\rho$=$\rho$$_{0}$exp(E$_{a}$/\textit{k}$_{B}$\textit{T}), where \textit{k}$_{B}$ is the Boltzmann constant. The activation energy E$_{a}$ was estimated to be 36 meV and 24 meV for \textit{x}=1.50 and 1.64 respectively. As the content of Fe increases, the resistivity at lower temperatures decreases rapidly. The preliminary indications of superconducting transitions are observed at 25.8 K and 27.6 K for \textit{x}=1.69 and 1.76, respectively, while no zero resistivity is observed above 2 K. In the \textit{x}=1.78 crystal, a sharp SC transition occurs at \textit{T}$_{c}$$^{mid}$=28.4 K and reaches zero resistivity at \textit{T}$_{c}$$^{zero}$=27.4 K. For the \textit{x}=1.83 sample, \textit{T}$_{c}$$^{mid}$ increases to 31.1 K. Although \textit{T}$_c$ of the sample with higher \textit{x} values, such as 1.84 and 1.85, does not increase much, and their volume faction of SC is quite larger (Fig.~\ref{Fig10}(d)).

\begin{figure}
  \includegraphics[width=8.3cm]{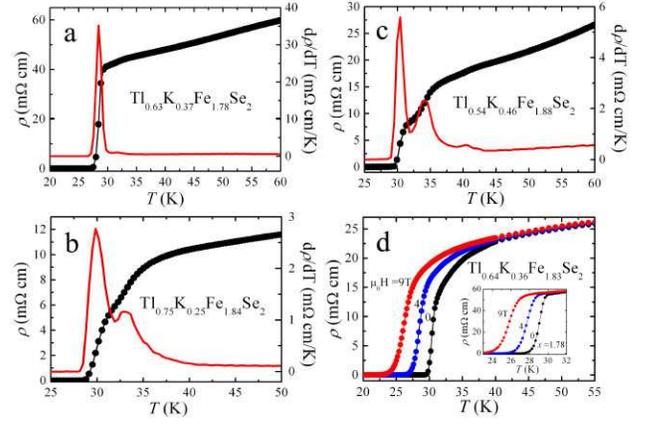}\\
  \caption{\label{Fig11}(color online) In-plane resistivity as a function of temperature for (a) Tl$_{0.63}$K$_{0.37}$Fe$_{1.78}$Se$_{2}$, (b) Tl$_{0.75}$K$_{0.25}$Fe$_{1.84}$Se$_{2}$, (c) Tl$_{0.54}$K$_{0.46}$Fe$_{1.88}$Se$_{2}$. (d)$\rho$(\textit{T}) of Tl$_{0.63}$K$_{0.37}$Fe$_{1.78}$Se$_{2}$ near the superconducting transition under different magnetic fields. From Ref. [26]}
\end{figure}

We observe two superconducing transitions in the crystal with \textit{x}=1.84,  \textit{T}$_{c}$=29.8 K and 33.0 K (Fig.~\ref{Fig11}(b)), in contrast with one transition in \textit{x}=1.78. In addition, the \textit{x}=1.88 sample with the highest Fe content exhibits three superconducting transitions with \textit{T}$_{c}$=30.4, 34.1 and 40.4 K (Fig.~\ref{Fig11}(c)). Resistivity measurements under a magnetic field up to 9 T confirmed the superconducting phase with \textit{T}$_{c}$=40 K for \textit{x}=1.83. This indicates that the domains with various Fe content values, corresponding to different carrier concentrations, likely exist in the crystals. In the major domain, superconducting phase with \textit{T}$_{c}$=30 K emerges. The volume fraction of the domains with higher \textit{T}$_{c}$ value and higher Fe content may be small because no diamagnetic signal appears at higher temperature. To our knowledge, this kind of phase separation was first observed in the Fe-based superconductors. Bulk SC with \textit{T}$_{c}$=40 K may emerge in the crystal with an optimal Fe-content, but it is difficult to realize in this system due to the Fe$^{2+}$ valence limitation. SC with \textit{T}$_{c}$=40 K emergence in the (Tl,K)Fe$_{x}$Se$_{2}$ crystals is reminiscent of the highest \textit{T}$_{c}$=38 K in the optimal doping Ba$_{1-x}$K$_x$Fe$_2$As$_2$ compound. It should be pointed out that Guo et al.\cite{xiaolong} reported on SC at 30 K in K$_x$Fe$_2$Se$_2$ compound almost at the same time.

\begin{figure}
  \includegraphics[width=8.3cm]{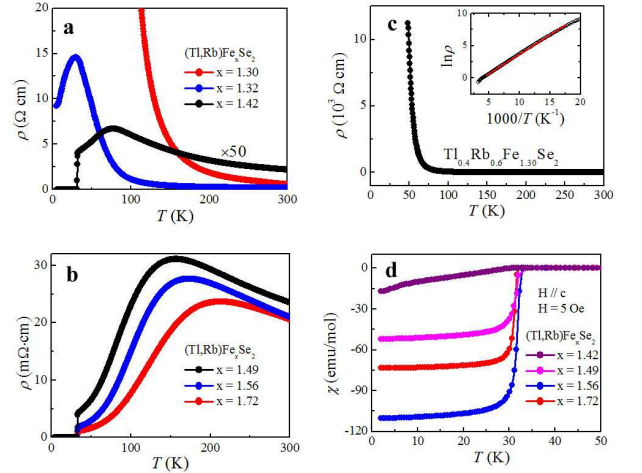}\\
  \caption{\label{Fig12}Temperature dependences of the in-plane resistivity of the (Tl,Rb)Fe$_x$Se$_2$ crystals for (a)\textit{x}=1.30, 1.32 and 1.42. (b)\textit{x}=1.49, 1.56 and 1.72. (c) \textit{x}=1.30. Inset: $\ln\rho$ as a function of 1/T. (d) magnetic susceptibility $\chi$(\textit{T}), measured at 5 Oe field with ZFC process, for the (Tl,Rb)Fe$_x$Se$_2$ (\textit{x} = 1.72, 1.56, 1.49 and 1.42) crystals. From Ref. [31]}
\end{figure}

At the same time, we attempted to grow the (Tl,Rb)Fe$_x$Se$_2$ single crystals by adding Rb in the starting materials. We have obtained a series of (Tl,Rb)Fe$_x$Se$_2$ (1.30 $\leq$\textit{x}$\leq$ 1.72) crystals. Figure~\ref{Fig12}(a)-12(c) show the plots of in-plane resistivity versus temperature, $\rho$(T). We first discuss the \textit{x}=1.30 sample, which is an insulator. As shown in Fig. 12(a), the $\rho$ at 300K is about 18.1 m$\Omega$$\cdot$cm. As \textit{T} decreases, $\rho$(T) increases rapidly and shows thermally activated behavior with an activation energy of \textit{E$_a$}$\sim$51.7 meV (the red fitting line in the inset of Fig~\ref{Fig12}(c)). With Fe content increasing, the bulk SC with \textit{T}$_c$=32 K emerges in the crystal, which is evidenced by the susceptibility measurements as shown in Fig.~\ref{Fig12}(d). The evolution from an insulator to a superconductor with the increase of Fe content has also been observed in the (Tl, Rb)Fe$_x$Se$_2$ system.

\begin{figure}
  \includegraphics[width=8cm]{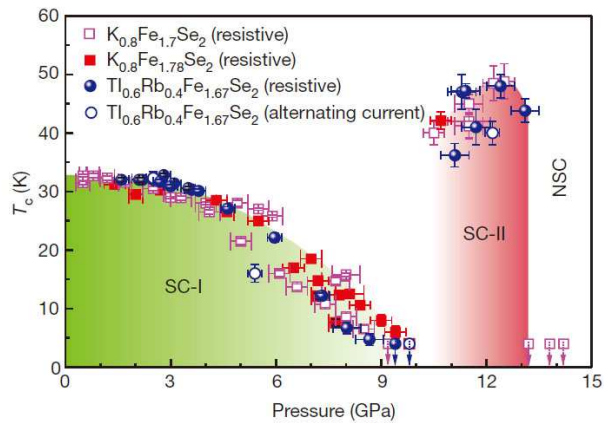}\\
  \caption{\label{Fig13}(color online) Pressure dependences of the \textit{T}$_{c}$ for the Tl$_{0.6}$Rb$_{0.4}$Fe$_{1.67}$Se$_{2}$, K$_{0.8}$Fe$_{1.7}$Se$_{2}$ and K$_{0.8}$Fe$_{1.78}$Se$_{2}$ single crystals. From Ref. [56]}
\end{figure}

Although SC with \textit{T}$_{c}$=40 K cannot be achieved by a traditional synthesis method under ambient pressure, SC with \textit{T}$_{c}$=48 K emerges under a higher pressure in the Tl$_{0.6}$Rb$_{0.4}$Fe$_{1.67}$Se$_{2}$, K$_{0.8}$Fe$_{1.7}$Se$_{2}$ and K$_{0.8}$Fe$_{1.78}$Se$_{2}$ samples \cite{Highpressure122} as shown in the phase diagram for them in Fig.~\ref{Fig13}. The original SC1 phase with \textit{T}$_{c}$ = 32 K under ambient pressure is rapidly suppressed to zero K between 9.2 and 9.8 GPa. At higher pressures, the SC2 phase appears, in which \textit{T}$_{c}$$\sim$48 K that is much higher than $T_C$ in SC1 phase. However, the pressure range of SC 2 is much narrower, unlike the usual parabolic pressure-tuning curve of \textit{T}$_{c}$. The SC suddenly disappears under 13.2 GPa.

Bulk superconductivity with \textit{T}$_c$=30K emerges in the TlFe$_x$Se$_2$ (\textit{x}$\geq$1.7), (Tl,K)Fe$_x$Se$_2$ (\textit{x}$\geq$1.72) and (Tl,Rb)Fe$_x$Se$_2$ (\textit{x}$\geq$1.42) crystals, separately. Naturally, a question arises: why the Fe content values of the crystals in which bulk SC starts to appear are different for these three systems. Now, we can understand this via a phase separation discussed below. Two phases always coexist in the SC crystals below the AFM transition temperature, \textit{T}$_N$. Fe sites are fully occupied in one phase, which should be responsible for bulk SC. While Fe vacancies form a $\sqrt{5}$ $\times$ $\sqrt{5}$ super-lattice in another phase, which should be an AFM insulator. Due to the difference in ionic radius between K and Rb, the phase separation emerges in the crystals with different Fe content values. The phase separation phenomenon is indeed observed in our single crystal XRD experiments.

\subsection{The Fe-vacancy super-lattice and AFM order in (Tl,K,Rb)Fe$_x$Se$_2$}

As mentioned above, we suggested \cite{Fang122} that there are two kinds of Fe-vacancy orders in the (Tl,K,Rb)Fe$_x$Se$_2$ system. One with the formula of (Tl,K,Rb)Fe$_{1.5}$Se$_2$, where the iron atoms have two or three iron neighbors, and the other with the formula of (Tl,K,Rb)Fe$_{1.6}$Se$_2$ (so called A$_2$Fe$_4$Se$_5$ phase), each iron atom has three iron neighbors. We pointed out that the insulating phase mentioned above should be associated with these Fe-vacancy orders. With these two kinds of Fe-vacancy orders, the band structure calculations indicate a band gap with the values of about 0.3-0.5 eV for TlFe$_{1.5}$Se$_2$ \cite{Xiaotaotheory1} and 60 meV for TlFe$_{1.6}$Se$_2$ \cite {CaoCTheory}. Of course, it is very important to determine which kind of Fe vacancy and AFM order exists most likely in this system.

Based on the neutron diffraction experiments, Bao et al.\cite {Baowei122} and Ye et al.\cite {YeFNeutron122}  concluded that the dominating phase had a composition of A$_{0.82}$Fe$_{1.62}$Se$_2$ (A=K, Tl/K and Tl/Rb) and Fe vacancies order in the $\sqrt{5}$ $\times$ $\sqrt{5}$ super-lattice below 460-580 K, corresponding to the second Fe-vacancy order structure suggested by us. At or slightly below this temperature, the spins align along the tetragonal \textit{c} axis in a block-checkerboard antiferromagnetic structure (BCAF-\textit{c}, where \textit{c} indicates the alignment of the spin). The four adjacent Fe ions, with a spin 3.31 $\mu$$_B$/Fe, form a square block with the spins aligned ferromagnetically. The total magnetic moment of the 4-iron-block is 13.4 $\mu$$_B$. This magnetic structure was later proved by Wang et al.\cite {PCDaiNeutron}. They concluded that this is a common crystalline and magnetic structure of the superconducting A$_2$Fe$_4$Se$_5$ (A=Tl, K, Rb and Cs)\cite {YeFNeutron122} compounds. The transmission electron microscopy (TEM) experiments \cite {LJQTME} have also confirmed this kind of Fe-vacancy order.

Very recently, May \textit{et al.}\cite{SpinabTl} reexamined the Fe vacancy and AFM order structure by means of single crystal neutron diffraction, nuclear forward scattering, and transmission electron microscopy for TlFe$_{1.6}$Se$_2$ crystal with complete chemical/vacancy order. They confirmed the existence of $\sqrt{5}$ $\times$ $\sqrt{5}$ Fe-vacancy super-lattice, and found that the Fe moment changes from along the \textit{c} direction into lying in the \textit{ab} plane below 100 K. And there may be two types of magnetic structures below 100K, one is a non-collinear, and the other is of an in-plane, block-checker antiferromagnetic structure. They argued that although the BCAF-\textit{c} structure is the main magnetic structure observed in the alkali-metal compounds, the TlFe$_x$Se$_2$ has a limited compositional window and the Tl sites is always fully occupied, there may appear additional magnetic phase transitions in the lower temperature, which is consistent with our susceptibility data \cite{Fang122} and the results reported in Ref.~\onlinecite{BCSalesTlFe1.6Se2}.

Recently, Zhao \textit{et al.} \cite {ZhaocollinearFeSe122} also re-checked the Fe vacancy super-lattice and magnetic structure in the insulating ($E_a$=500 meV) and semiconducting ($E_a$=40 meV) K$_y$Fe$_x$Se$_2$ crystals. They found that there is only a $\sqrt{5}$ $\times$ $\sqrt{5}$ Fe-vacancy super-lattice corresponding to the BCAF-\textit{c} magnetic structure, in the insulating K$_{0.8}$Fe$_{1.6}$Se$_2$ crystal. However, a $\sqrt{2}$ $\times$ $\sqrt{2}$ Fe vacancy super-lattice is also observed in the semiconducting K$_y$Fe$_x$Se$_2$ crystal in addition to the $\sqrt{5}$ $\times$ $\sqrt{5}$ super-lattice. This new Fe vacancy order in which iron atoms have two or three iron neighbors is the same as that discussed in TlFe$_{1.5}$Se$_2$ \cite{Fang122} as shown in Fig.~\ref{Fig7}(c). According to Zhao's neutron diffraction experiments, the magnetic structure in this Fe vacancy super-lattice has a stripe type AFM order with N\'{e}el temperature 280 K, which is the same as that in other iron pnictides. They concluded that this AFM semiconductor is the parent compound of the supercondicting AFe$_x$Se$_2$ compounds, interpolating between the antiferromagntic Mott insulator of the cuprates and the antiferromagnetic semimetal of the iron pnictides. Importantly, it indicates that the correlation physics plays an important role in the magnetism for this system, which is consistent with the expectation of our work \cite{Fang122}. Of course, there are still debates about which antiferromagnetic order is the parent of superconductivity emerging in this system.

\subsection{Phase separation in (Tl,K,Rb)Fe$_x$Se$_2$ system}

\begin{figure}
\includegraphics[width=8.5cm]{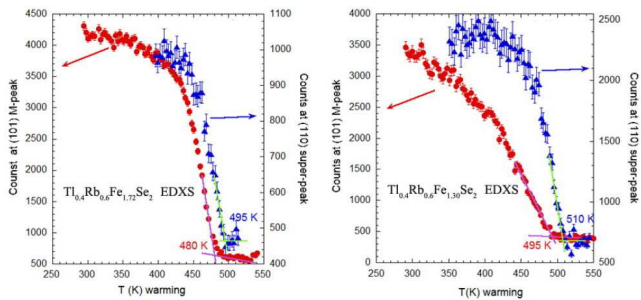}\\
\caption{\label{Fig14}Temperature dependences of the intensity of (110) peak, corresponding to $\sqrt{5}$ $\times$ $\sqrt{5}$ super-lattice, and the intensity of (101) peak, corresponding to BCAF-\textit{c} lattice, in the powder neutron diffraction pattern. The left panel is for the superconducting Tl$_{0.4}$Rb$_{0.6}$Fe$_{1.72}$Se$_2$ crystals, and the Right panel is for the insulator Tl$_{0.4}$Rb$_{0.6}$Fe$_{1.30}$Se$_2$ crystals. }
\end{figure}

\begin{table} 
\centering
\parbox{8.5cm}{\caption{\label{tab:table1}The fractions of I4/\textit{m} and I4/\textit{mmm} phases obtained by fitting to ND data measured at different temperatures for superconducting Tl$_{0.374}$Rb$_{0.374}$Fe$_{1.764}$Se$_{2}$ crystals.}}
\begin{tabular}{p{1.5cm}<{\centering} p{1.5cm}<{\centering} p{1.5cm}<{\centering} p{1.5cm}<{\centering} p{1.5cm}<{\centering}}
\hline
\hline
 & 30K & 300K & 400K & 550K \\ \hline
 I4/\textit{m} & 70.8\% & 70.9\% & 70.0\% & 0\% \\
 I4/\textit{mmm} & 20.8\% & 21.5\% & 21.8\% & 91.56\% \\
\hline
\hline
\end{tabular}
\end{table}

\begin{table*} 
\centering
\parbox{13.5cm}{\caption{\label{tab:table2}Refined structure parameters for the I4/\textit{mmm} phase, obtained by fitting to ND data measured at 30, 300, 400 and 550 K for superconducting Tl$_{0.374}$Rb$_{0.374}$Fe$_{1.764}$Se$_{2}$. At 550K, Tl/Rb at 2\textit{a} (0 0 0), Fe 4\textit{d} (0 1/2 1/4) and Se at 4\textit{e} (0 0 \textit{z}) sites. 3.66\% Fe and 4.78\% FeSe as impurity are taken into account in the refinement.}}
\begin{tabular}{p{1cm}<{\centering} p{1cm}<{\centering} p{2.3cm}<{\centering} p{2.3cm}<{\centering} p{2.3cm}<{\centering} p{2.3cm}<{\centering}}
\hline
\hline
 & & 30K & 300K & 400K & 550K \\ \hline
 & \textit{a}({\AA}) &	3.8264(3) &	3.8423(3) & 3.8550(4) & 3.9226(1) \\
 & \textit{c}({\AA}) & 14.589(3) &	14.820(2) &	14.894(4) &	14.4414(9) \\
Rb/Tl &	\textbf{n} & \textbf{0.268(9)} &	\textbf{0.267(8)} &	\textbf{0.26(1)} & \textbf{0.374(5)}\\
Fe & \textbf{n} &	\textbf{1} &	\textbf{1} &	\textbf{1} &	\textbf{0.88(1)}	 \\
Se & \textbf{n} &	\textbf{1} &	\textbf{1} &	\textbf{1} & \textbf{1} \\
& \textit{z} &	0.3503(5) &	0.3515(4) &	0.3543(7) & 0.3532(2) \\
\hline
\hline
\end{tabular}
\end{table*}

\begin{table*} 
\centering
\parbox{14.2cm}{\caption{\label{tab:table3}Refined structure parameters for the I4/\textit{m} phase, obtained by fitting to ND data measured at below \textit{T}$_N$ for superconducting Tl$_{0.374}$Rb$_{0.374}$Fe$_{1.764}$Se$_{2}$.}}
\begin{tabular}{p{1cm}<{\centering} p{1.8cm}<{\centering} p{2.3cm}<{\centering} p{2.3cm}<{\centering} p{2.3cm}<{\centering} p{2.3cm}<{\centering}}
\hline
\hline
 & & 30K & 300K & 400K & 500K \\ \hline
 & \textit{a}({\AA}) &	8.6685(4) &	8.7067(4) &	8.7303(5) &	8.7487(5) \\
 & \textit{c}({\AA}) & 14.235(1) &	14.335(1) &	14.355(2) &	14.405(2) \\
Rb1/Tl1 &	\textbf{n} & \textbf{0.408(6)} &	\textbf{0.409(5)} &	\textbf{0.37} &	\textbf{0.37}\\
Rb2/Tl2 &	\textbf{n} & \textbf{0.408(6)} &	\textbf{0.409(5)} &	\textbf{0.37} &	\textbf{0.37}\\
& \textit{x} &	0.399(2) &	0.396(1) &	0.394(2) &	0.396(3) \\
& \textit{y} &	0.190(2) &	0.194(2) &	0.181(2) &	0.184(3) \\
Fe1 & \textbf{n} &	\textbf{0.15(2)} &	\textbf{0.10(1)} &	\textbf{0.06(2)} &	\textsc{0.22(2)} \\
Fe2 & \textbf{n} &	\textbf{0.92(1)} &	\textbf{0.943(8)} &	\textbf{1} &	\textbf{1} \\
& \textit{x} &	0.1992(8) &	0.1996(6) &	0.1955(9) &	0.198(1) \\
& \textit{y} &	0.0909(6) &	0.0895(4) &	0.0910(7) &	0.0930(8) \\
& \textit{z} &	0.2491(9) &	0.2477(9) &	0.252(1) &	0.252(1) \\
Se1 & \textbf{n} &	\textbf{1} &	\textbf{1} &	\textbf{1} &	\textbf{1} \\
& \textit{z} &	0.1291(9) &	0.1333(8) &	0.132(1) &	0.133(2) \\
Se2 & \textbf{n} &	\textbf{1} &	\textbf{1} &	\textbf{1} &	\textbf{1} \\
& \textit{x} &	0.1077(9) &	0.1080(7) &	0.107(1) &	0.111(1) \\
& \textit{y} &	0.288(1) &	0.2946(10) &	0.294(1) &	0.298(2) \\
& \textit{z} &	0.1480(3) &	0.1483(2) &	0.1490(4) &	0.1498(4) \\
& \textit{Rp} (\%)	& 5.50 &	5.45 &	6.08 &	7.23\\
& \textit{Rwp} (\%)	& 7.01 &	6.75 &	7.67 &	9.32\\
& \textit{$\chi$}$^{2}$	& 2.399 &	1.785 &	1.384 &	1.985\\
\hline
\hline
\end{tabular}
\end{table*}

Another important issue is the coexistence between SC and magnetism in the (Tl,K,Rb)$_y$Fe$_x$Se$_2$ system. A lot of techniques have been employed to understand this interesting property. Now, the coexistence between SC and magnetism is believed to be due to the fine-scale phase separation, with each property being associated with a different composition and/or degree of order.\cite {PSXRD,PSMOSS1,PSOPTICAL,NMRML,CharnukhaPS,PSMAG} All observations via transmission electron microscopy (TEM) \cite {LJQTME,PSTEM2}, nano-focused x-ray diffraction \cite {PSXRD} and M\"{o}ssbauer spectroscopy \cite {PSMOSS1,PSMO2,Mossbauerour} have confirmed the existence of the phase separation over length scales of 10-100 nm in the superconducting samples. Although superconducting samples with single phase (structural and chemically homogenous) have not been produced up to now, it is not hard to grow the insulating single phase crystal with BCAF-\textit{c} magnetic structure.

We chose two kinds of (Tl,Rb)Fe$_x$Se$_2$ crystals to determine their crystal and magnetic structure by neutron diffraction experiments. The first one is an insulator as shown in Fig.~\ref{Fig12}(c), \textit{i.e.}, Tl$_{0.4}$Rb$_{0.6}$Fe$_{1.30}$Se$_2$. Above 510 K, the powder neutron diffraction (ND) data can be well fitted by a pure 122 structure phase with the I4/\textit{mmm} space group, and the composition is determined to be Tl$_{0.452}$Rb$_{0.452}$Fe$_{1.578}$Se$_2$. Below 510 K, the ND data can be well fitted by a pure 245 block phase with I4/\textit{m}. At 495K, a BCAF-\textit{c} magnetic structural transition occurs. Temperature dependences of the intensities of the (110) peak corresponding to the $\sqrt{5}$ $\times$ $\sqrt{5}$ super-lattice and the (101) peak corresponding to the BCAF-\textit{c} lattice are shown in the right of Fig.~\ref{Fig14}.

The second sample is a superconductor as shown in Fig.~\ref{Fig12}(b), \textit{i.e.}, Tl$_{0.4}$Rb$_{0.6}$Fe$_{1.72}$Se$_2$. Above 495 K, the powder neutron diffraction (ND) data can also be well fitted by a pure 122 phase with the I4/\textit{mmm} space group and a composition Tl$_{0.374}$Rb$_{0.374}$Fe$_{1.764}$Se$_2$. But below 495 K, two phases both emerge in these crystals. One phase is (Tl,Rb)$_{0.818}$Fe$_{1.563}$Se$_2$ with the I4/\textit{m} space group, which becomes BCAF-\textit{c} below 480K shown in Fig.~\ref{Fig14}(Left), its volume fraction is about 70$\%$, and its refined structure parameters from ND data below \textit{T}$_N$ are shown in Table 3. The other phase is (Tl,Rb)$_{0.532}$Fe$_2$Se$_2$ with the I4/\textit{mmm} space group, and its refined structure parameters from ND data below \textit{T}$_N$ are shown in Table 2. No magnetic order is discovered in this phase and its volume fraction is about 22$\%$. Little change is found in the relative ratio between two phases respectively at 490K and 30K as listed in Table 1. We believe that (Tl,Rb)$_{0.534}$Fe$_2$Se$_2$ phase is responsible for the superconductivity in this compound.

\section{Conclusions}

In this paper, we briefly discuss the discovery of SCs, structural and magnetic structures in the iron-chalcogenide compounds. Structurally, the layer consisting of the edge-sharing tetragonal FeX$_4$ (X=Se, Te, S) without Fe vacancies is considered to be responsible for the SCs in Fe(Te,Se,S) and (Tl,K,Rb)Fe$_x$Se$_2$ compounds. However, the excess Fe atoms at the interstitial sites between the adjacent Fe$_2$X$_2$ layers provide a pair-breaking effect and is antagonistic to SC; in contrast, the superconducting phase in (Tl,K,Rb)Fe$_x$Se$_2$ system cannot exist in the absence of the insulating phase with the iron vacancies, indicating an intimate correlation between these two separated phases. Magnetically, the parent compound of Fe(Te,Se,S) system, FeTe, is in the (incommensurate)commensurate AFM order, depending on the excess Fe, whereas this AFM order can be suppressed by isovalent doping rather than applying physical pressure, \cite{FeTeP1,FeTeP2} and survive as a short-range magnetic correlation. Although there are discrepancies on the phase diagram of the Fe(Te,Se,S) system, we have confirmed that bulk SC with ($\frac{1}{2}$, $\frac{1}{2}$) spin resonance can coexist with a long-range AFM order with the wave vector \textbf{Q}$_{AFM}$=($\frac{1}{2}$, 0). So there may be a common feature among the phase diagrams for all Fe-based superconductors. As for the (Tl,K,Rb)Fe$_x$Se$_2$ system, so called '245' phase with $\sqrt{5}\times\sqrt{5}$ super-lattice, it is in a block AFM state, while the superconducing phase with no Fe vacancies is non-magnetic. Which phase is the parent compound of this newly found (Tl,K,Rb)Fe$_x$Se$_2$ superconductor is still in hot debate, further investigation is needed.

\begin{acknowledgments}
Project supported by the National Basic Research Program of China under (Grant Nos. 2011CBA00103, 2012CB821404, and 2009CB929104), the National Natural Science Foundation of China (Grant Nos. 10974175, 10934005, and 11204059), the Zhejiang Provincial Natural Science Foundation of China (Grant No. Q12A040038), and the Fundamental Research Funds for the Central Universities of China.
\end{acknowledgments}

\end{document}